# Local Convergence and Global Diversity: The Robustness of Cultural Homophily

**Andreas Flache*, Michael W. Macy[+]**

***Department of Sociology, University of Groningen, a.flache@rug.nl**

**[+]Department of Sociology, Cornell University, mwm14@cornell.edu**

**DRAFT VERSION. DO NOT QUOTE WITHOUT AUTHORS' PERMISSION.**

## 1. Introduction

Cultural diversity is both persistent and precarious. People in different regions of the world are increasingly exposed to global influences from mass media, internet communication, interregional migration and mass tourism. English is rapidly becoming Earth's *Lingua Franca*, and Western music, films, and fashions can be found even in remote villages around the globe. While some authors point to the pervasiveness of cultural differences, others expect globalization to lead to the emergence of a universal monoculture (cf. Greig, 2002). This expectation is grounded not only in empirical observation of recent trends (Friedman, 2005) but also in formal theoretical models of social influence that predict an inexorable tendency toward cultural convergence (French 1956; Abelson 1964). Abelson proved that convergence to global consensus was inevitable in what he called a "compact" network – in which there are no subgroups that are entirely cut off from outside influences. Friedkin and Johnson (1999) showed later that even in compact networks some diversity may remain if actors' initial opinions reflect persistent individual interests that are entirely neutralized by outside influences.

Recently, formal theories of social influences have pointed to homophily as a new explanation of cultural diversity. Building on previous work by Carley (1991) and by Nowak et al (1990), Axelrod (1997; see also Mark 1998) proposed an elegant extension of social influence models that allows the network to change in response to changes in the distribution of cultural attributes. Like earlier formal models, Axelrod assumed one of the fundamental prin-ciples of human interaction – social influence, or the tendency to become more similar to influential others (Festinger et al, 1950). However, unlike earlier social influence models that assumed fixed social networks, Axelrod incorporated the equally fundamental principle of "likes attract", or homophily (McPherson et al, 2001). Combined with social influence, homophily generates a self-reinforcing dynamic in which growing similarity strengthens attraction, attraction increases influence, and influence leads to greater similarity.

This circular dynamic might appear to merely strengthen the tendency towards global convergence demonstrated by models with a fixed network structure (e.g. Abelson, 1964). However, Axelrod's computational studies showed how local convergence can preserve global diversity. The key assumption is cultural speciation. Axelrod assumed social interaction becomes impossible between actors who have nothing in common, analogous to the inability of sexual organisms with a common ancestor to mate once they differentiate beyond a critical threshold. Once influence between regions becomes impossible, their cultures evolve along divergent paths.

A range of follow up studies supported Axelrod's basic conclusions (e.g. Mark 1998, 2003; Parisi et al 2003), albeit with some modifications (e.g. Shibanai et al 2001; Greig, 2002). However, Klemm et al (2003a,b) challenged the robustness of stable diversity in Axelrod's model. They showed that a population that exhibits stable diversity under Axelrod's assumptions "drifts" toward monoculture in the presence of even very small amounts of random "cultural mutation." Mutation increases local heterogeneity by allowing agents to spontaneously deviate from their neighbors by altering one of their cultural traits. Paradoxically, this heterogeneity promotes not diversity but global convergence. Local convergence can trap a population in an equilibrium in which influence is no longer possible because neighbors are either identical or totally different. Mutation disturbs this equilibrium, allowing influence to proceed to a new equilibrium with less diversity than the last, and so on, until no differences remain. Given that in most cultural influence processes there are at least a few deviants who spontaneously introduce cultural mutations to their neighbors we are thus back to our original question: *How is stable diversity possible*?

We propose that Axelrod is right about homophily, but his model was too conservative in its implementation. This becomes plausible when we notice two hidden assumptions in Axelrod's model that attenuate the effects of homophily on the formation of influential ties. First, Axelrod has an implicit threshold of similarity below which agents cannot influence one another. He made the highly conservative assumption that this threshold was at the theoretical lower limit of zero influence. Klemm et al. also assumed a zero threshold of similarity. At this level, there will be some positive probability of influence except in the special case that two agents have absolutely nothing in common. As the number of cultural features increases, the probability that two randomly chosen agents will have nothing in common approaches zero, in which case, every agent has some attraction to every other agent. Raising the influence threshold strengthens homophily by limiting attraction to a subset of neighbors whose similarity exceeds a critical value.

Second, Axelrod assumed that influence was dyadic. Dyadic influence attenuates the effects of homophily by allowing agents to imitate a deviant in isolation from the countervailing influence from all the other conforming neighbors. That is, Axelrod assumes that agents focus on only one neighbor at a time, ignoring the traits of all their other neighbors. This allows agents to imitate a deviant neighbor while ignoring the influence of conformist neighbors. Instead, we allowed a randomly chosen agent to be simultaneously influenced by multiple neighbors, rather than by just one randomly chosen neighbor at a time (cf. Parisi 2003 for a similar approach). By allowing multiplex influence, it is no longer possible for a deviant to lure its neighbors by influencing them one at a time. This strengthens the effects of homophily by insuring that agents can never be influenced in a direction that leaves them with less in common with their neighbors overall. If homophily is the mechanism by which global convergence generates local diversity, then strengthening the tendency toward convergence might have the counterintuitive effect of allowing stable diversity to emerge.

In the section that follows, we describe our extensions of the Axelrod model designed to strengthen homophily. Section 3 presents results of computational experiments. In section 4, we discuss the our analysis and identify promising directions for future research.

## 2. An Extended Model of Cultural Influence

We began by replicating Axelrod's (1997) dynamic network model of homophily and influence, with no mutation, a zero threshold of similarity, and dyadic influence (from only one neighbor at a time). We then extended this model by introducing the assumption of mutation that causes diversity to collapse, and two additional assumptions – higher thresholds and multiplex influence – that strengthen homophily. This allows us to test the hypothesis that the fragility of cultural diversity in the original Axelrod specification reflects too little homophily, not too much. If this hypothesis is confirmed, it means that Axelrod did not go far enough. Not only does global convergence produce local diversity, but the stronger the convergent tendency, the more stable the emergent diversity.

In the original Axelrod model, the population consists of *N* agents distributed over a regular bounded (non-toroidal) lattice, where each cell is occupied by a single agent who can interact only with the four neighbors to the N, S, E, and W (a von Neumann neighborhood). At any point in time, the cultural profile for an agent is a vector of *F* features[1]. On any feature

1 We follow Axelrod in referring to cultural dimensions – like religion, musical preference, ethnicity, or occupation – as "features," and locations on these dimensions – like Catholic, Baptist, or Muslim – as "traits."

$f$, an agent has a trait $q$ represented by an integer value in the range of $q = \{0...Q - 1\}$, where $Q$ is the number of possible traits on that feature. Formally, the cultural profile $C$ of agent $i$ is

$$C_i = (q_{i1}, q_{i2}, ..., q_{iF}), \quad q_{ix} \in \{0,1,..., Q-1\} \subset N_0. \qquad (1)$$

Following Axelrod, agents in our baseline model start out with a random cultural profile in which each feature is assigned a trait $q$ from a uniform distribution. In every discreet time step $t$, an agent $i$ is randomly chosen from the population, and $i$'s cultural profile is updated through interaction with a randomly chosen neighbor $j$. The probability $p_{ij}$ that $j$ will interact with $i$ is given by their cultural overlap, defined as the proportion of all $F$ features with identical traits. Two agents $i$ and $j$ have identical traits on feature $f$ if $q_{if} = q_{jf}$. If $i$ and $j$ do not interact, another agent is randomly chosen for possible interaction with a randomly chosen neighbor. If $i$ and $j$ do interact, $j$ influences $i$ by causing $i$ to adopt $j$'s trait on a feature randomly chosen from the set of features on which $q_{if} \neq q_{jf}$.

Up to this point, our specification of the influence process is identical to the influence process assumed by Axelrod (1997) and also by Klemm et al (2003a,b).

We introduced three changes to this "baseline" model. The first change – cultural mutation – is expected to destabilize diversity, leading to the inevitability of monoculture. The second two changes – higher thresholds of similarity and simultaneous influence from multiple neighbors – are hypothesized as antidotes that allow stable diversity to emerge even under conditions that would otherwise guarantee monoculture.

1. *Cultural mutation.* Following Klemm et al. (2003a,b), we implemented cultural mutation as a small probability $r$ that a randomly chosen trait will be perturbed to a new value randomly chosen over the interval $\{0\ldots Q - 1\}$.

2. *Thresholds of similarity.* Our other two extensions of Axelrod's model are intended to strengthen the effects of homophily by precluding influence from neighbors who are insufficiently similar. Raising the threshold of similarity relaxes the linear relationship that Ax-elrod and Klemm et al. assume between influence and overlap. The linear form carries the implicit assumption that the threshold of similarity (below which influence is precluded) is at the theoretical lower limit, which means that any shared traits, no matter how few, are sufficient for there to be a positive probability of influence. We simply raised this lower limit by introducing a threshold $\tau$ ($0 \leq \tau \leq 1$) above which the probability of influence remains linear with cultural overlap (identical to Axelrod) but below which the probability is zero:

$$p_{ij} = \begin{cases} 0 & \text{if } o_{ij} \leq \tau \\ o_{ij} & \text{otherwise} \end{cases} \qquad (2)$$

where $o_{ij}$ represents the extent of cultural overlap given by the proportion of features with identical traits (cf. Deffuant et al, 2005, for a similar approach called “bounded confidence”).

3. *Multiplex influence.* In the baseline model, an agent is randomly chosen from the population who then randomly picks a single neighbor to be imitated with a probability corresponding to the proportion of similar features. In order to allow multiplex influence, we assume that all neighbors can potentially be included in the set of influential neighbors, each with a probability corresponding to the proportion of similar features. Once the set of influential neighbors has been formed, the focal agent adopts on a randomly selected feature the trait that is supported by the largest number of influential neighbors. In case a tie occurs, a random choice is made between all maximally popular traits. This preserves the original dyadic influence model as a condition nested within a more general form, such that multiplex influence reduces to the original model when an agent has only one influential neighbor. With more than one influential neighbor, it is no longer possible for a deviant to induce a neighbor to adopt a trait that is an outlier within the distribution over all influential neighbors.

More precisely, we assume that, at a given timepoint *t*, *i* can be simultaneously influenced by a subset of *i*’s neighbors, where each neighbor *j* is selected into the subset of influential neighbors $I_i$ with a probability corresponding to the cultural overlap between *i* and *j*, as specified in equation (4). Following Axelrod, a random number $r_{ij}$ is drawn from a uniform distribution, where $0 \leq r_i \leq 1$, and this value is then compared with the overlap $o_{ij}$ between *i* and *j*. If $o_{ij}$ exceeds $r_{ij}$, then *j* is added to *I*:

$$I_i = \left\{ j \in J(i) \,\middle|\, o_{ij} \geq r_{ij} \right\}, \qquad (3)$$

where *J*(*i*) denotes the set of all neighbors *j* of *i*.

Agent *i* then adopts the modal trait on *f* within $I_i$. That is, *i* observes the trait $q_{fk}$ on *f* of every member *k* of the influence set. The modal trait *m*, is obtained as the number of influential neighbors of *i*, $v_{fm}$, who have trait *m* on feature *f*, or

$$\forall m \in [0..Q-1]: v_{fm} = \left\| \{ k \in I_i \,|\, q_{fk} = m \} \right\| \qquad (4)$$

If more than one trait satisfies this criterion, *i* chooses randomly from among those traits.

Importantly, our extended model reduces to Axelrod's original specification if the rate of cultural mutation is zero, the threshold of similarity is zero, and there is only one influential neighbor, that is, $\tau = r = 0$, and $k$=1. If $r$>0, *i* has a positive probability of spontaneously changing one of its cultural traits. If $\tau$>0, cultural overlap has no effect on interaction until the amount of overlap exceeds the critical value. Finally, if influence is multiplex, an agent adopts the modal trait among all influential neighbors.

## 3. Results

### 3.1 Baseline model

As a baseline, we replicated Axelrod's (1997) model, using the standard scenario that Axelrod employed throughout most of his simulations: a 10x10 grid (*N*=100), five features (*F*=5), 15 traits per feature (*Q*=15), no mutation ($r$ = 0), and a zero threshold of similarity ($\tau$ = 0). Using

Axelrod's measure of diversity as the number of cultural regions in equilibrium,[2] we obtained results identical to those reported by Axelrod. From a random start, cultural diversity rapidly declines but the rate of decline slows exponentially, culminating in an equilibrium at which influence is no longer possible because the overlap between all pairs of neighbors is either 0 or 1. Based on 10 realizations, each from a random start, Axelrod reported for this condition an average of 20 stable cultural regions in equilibrium. We obtained an average of 19.55 based on 100 realizations (and we suspect that Axelrod simply rounded this number up to 20). As an additional validity test, we replicated all other results reported in Axelrod (1997) for the effects of network size $N$, neighborhood size $k$, and the number of features $F$ and traits $Q$.

### 3.2 Diversity Collapses with Cultural Mutation

We next replicated Klemm et al. (2003a) by introducing the possibility for cultural mutation ($r > 0$). All other parameters were identical to the baseline model described above. For reliability, we ran 20 realizations of the experiment, allowing each run to continue until the level of diversity became stable or the number of iterations exceeded $10^6$. Our results were again identical to what previous work found: Low rates of mutation ($r \lesssim 10^{-3}$) were sufficient to destabilize diversity so that the system gradually moves from a random start towards a metastable state close to cultural homogeneity.

Figure 1 reports the typical dynamics for a mutation level of $r = 10^{-3}$. Over 20 realizations, we averaged 1.55 cultural regions in the end-state, compared to 20 regions with $r = 0$. Panel 3 of Figure 1 also illustrates how cultural diversity no longer declines monotonically in the presence of mutation. This shows that mutations have two paradoxical effects. On the one hand, mutations create bridges for social influence between otherwise isolated cultural re-gions. These bridges act as an annealing mechanism that allows the system to escape a local equilibrium and find a new equilibrium with less diversity. On the other hand, mutation can also spark the spontaneous emergence of new cultural regions when new traits appear and spread over the network. At the low level of mutation used for Figure 1, the first effect prevails. Mutations prevent the population from "freezing" in multicultural states with multiple disconnected cultural regions, leading inevitably to a monoculture that is highly resistant to further disturbance. As Klemm et al showed, at this point mutations may cause spontaneous "fads" that lead from one monoculture into another, but the level of diversity remains consistently at its minimum except for short transition periods.

[2] Klemm et al (2003a,b) use the relative size of the largest cultural region in equilibrium.

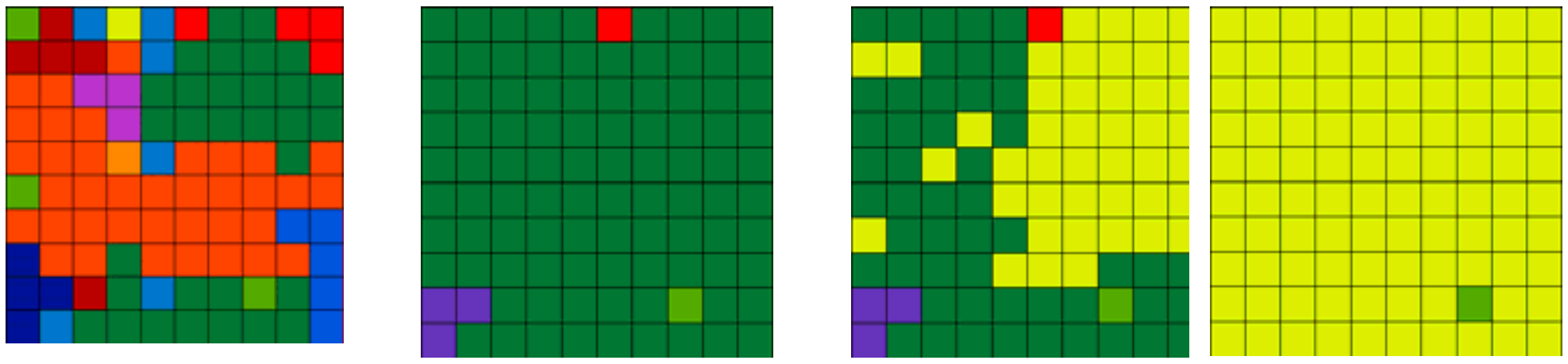

*t*=60K *t*=210K *t*=350K t=390K

***Figure 1a*. Representative dynamics with mutation (*N*=100, *k*=4, *F*=5, *Q*=15, $\tau = 0$, *r*=0.001, dyadic interaction). Color of a cell represents trait on feature 1.**

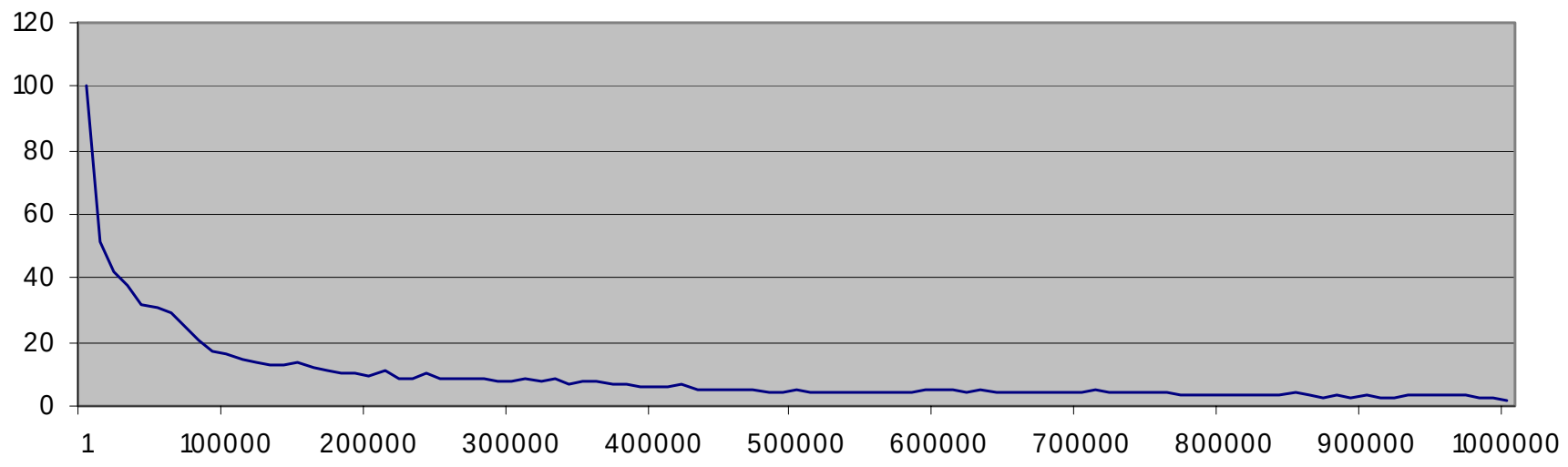

# iterations

***Figure 1b*. Change of average number of cultural regions with mutation (*N*=100, *k*=4, *F*= 5, *Q*=15, $\tau = 0$, *r*=0.001, dyadic interaction), based on 20 replications.**

Our results confirm the fragility of Axelrod's results reported by Klemm et al. Very small amounts of cultural mutation are sufficient to cause stable diversity to collapse. In the next sections, we report results that show how interaction thresholds and multiplex interaction can restore the robustness of Axelrod's results.

### 3.3 Higher thresholds of similarity

We begin by considering the threshold effect. Axelrod and Klemm et al. assumed a threshold at the theoretical lower limit of zero, which means neighbors always have some positive probability to imitate one another, except in the unlikely case that they do not have a single shared trait. This appears to be an appropriately conservative assumption when testing for the collapse of diversity. However, we need to know if the effects of mutation and metric features depend on the assumption that the threshold is zero.

To test this, we replicated the studies of Klemm et al. except that we raised the threshold of similarity from 0 to 1 in steps of 0.05. For reliability we report averages across 20 independent replications per condition. Outcomes were measured when equilibrium was approached. We stopped the simulation when there was no more change of diversity within the last 200.000 iterations, or the number of iterations exceeded $10^7$. Figure 2 shows the number of metastable cultural regions broken down by the rate of mutation and the threshold level.

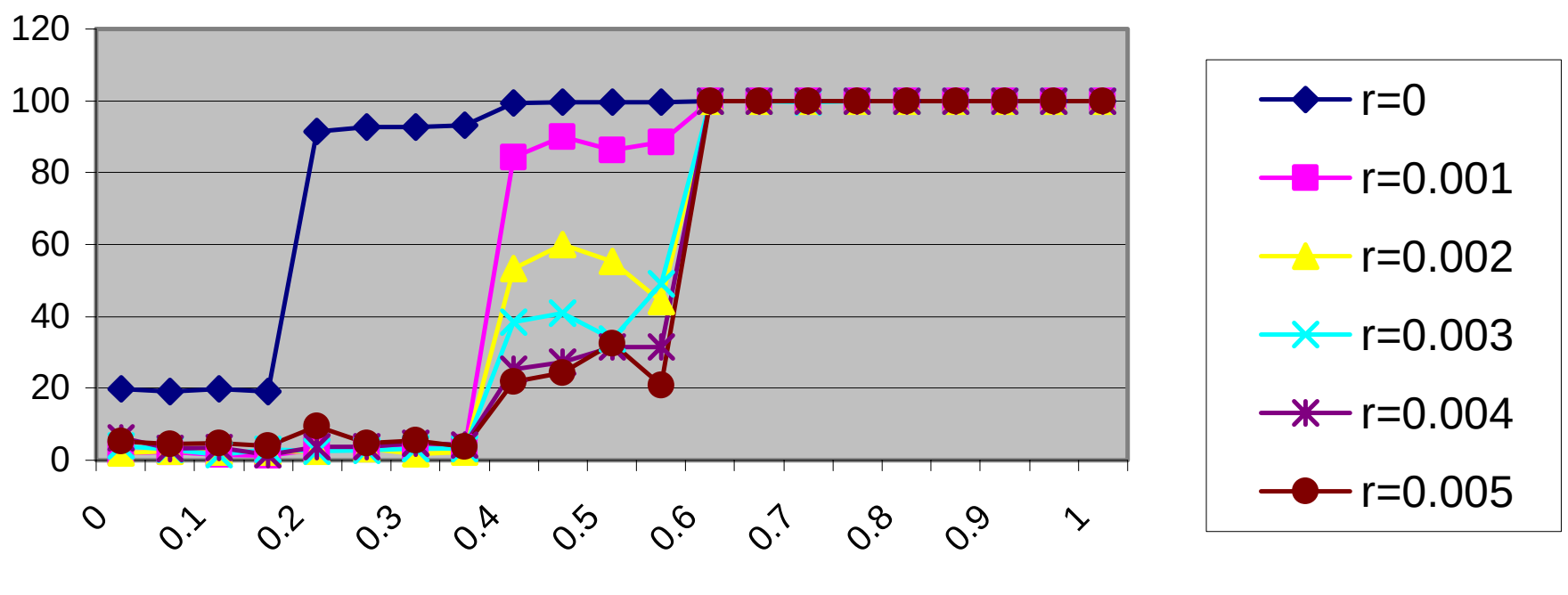

***Figure 2*. Effect of threshold $\tau$ on number of cultural regions in equilibrium for five levels of mutation, with $F$=$F_n$ = 5, $Q$ = 15, dyadic interaction. Averages based on 20 replications per condition after maximally $10^7$ iterations.**

The results reported in Figure 2 clearly reveal the paradoxical effects of mutation. Figure 2 shows that cultural diversity can be sustained, even with mutation, so long as the threshold is at least 0.4. As $\tau$ increases further, the number of cultural regions in equilibrium approaches the theoretical maximum of 100, indicating a population in which almost all agents are cultural isolates. Inspection of the convergence times until equilibrium shows why. From a random start, thresholds above 0.4 are sufficient to effectively preclude interaction. It becomes extremely unlikely that two neighboring agents will have sufficient cultural overlap to obtain a positive probability of influence or a non-zero weight.

Figure 2 also shows how this pattern changes as thresholds and mutation rates increase. With a threshold of 0.2 or lower, the number of cultural regions without mutation is about 20, consistently with what Axelrod found without thresholds. But with mutation, diversity can only be sustained when the interaction threshold exceeds the critical level of $\tau=0.4$. When thresholds drop below 0.4 (corresponding to two or one common trait in a model with $F=5$) does diversity collapse into monoculture under mutation. However, as soon as the threshold reaches 0.6, the level of diversity approaches its theoretical maximum of 100 different cultural regions for the mutation rates that we simulated in this experiment. Further analyses showed that these effects did not change qualitatively even with at least $10^9$ iterations.

Sufficiently high thresholds resurrect diversity in the face of cultural mutation, but does the extended model generate the same substantive implications than Axelrod's original model of cultural dissemination? We found at least one notable exception. Axelrod (1997) argued that a larger number of cultural features dramatically reduces cultural diversity. We found that the underlying mechanism depends critically on the assumption of zero thresholds. From a random start, any additional feature increases the chances that neighbors have identical traits larger than zero on at least one feature. With a threshold of zero, this implies that more features reduce the chances of cultural isolation and thus also the expected level of sta-ble cultural diversity. However, with higher thresholds a “critical mass” of identical features is needed before neighbors are sufficiently similar to have a chance to interact. This may turn Axelrod’s effect upside down: the more features, the larger this critical mass and thus the smaller may be the chances that identical traits occur on the required number of features. To test this intuition, we returned to Axelrod’s assumptions of zero mutation ($r$=0) and increased the number of features $F$ between 1 and 50 for thresholds ranging from $\tau = 0$ to $\tau = .2$ in steps of 0.05. Figure 3 shows the results.

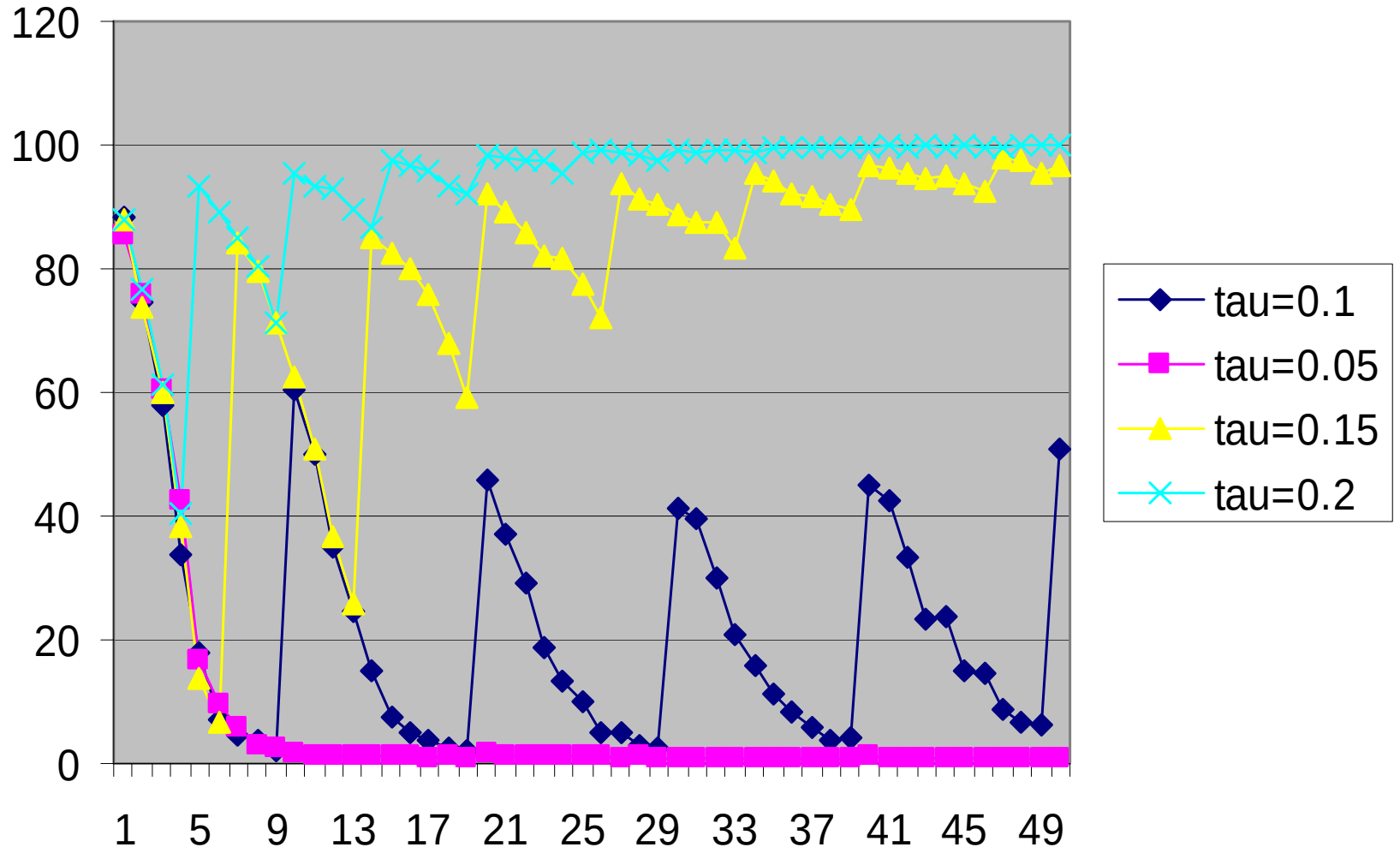

***Figure 3*. Effect of number of number of features (*F*) on number of cultural regions in equilibrium at four different levels of $\tau$ (*F*=1..20 , *Q*=15, *r* = 0, dyadic influence). Averages based on 20 replications per condition after convergence.**

Figure 3 shows that as $\tau$ approaches zero, the effect of $F$ is similar to what Axelrod reported for $\tau$=0. However, as $\tau$ increases, the effect of $F$ reverses, leading to a dramatic secular increase in diversity. Note also the wave-shaped pattern. This is an artifact of the intervals of $F$ within which the critical number of features needed for influence remains constant, such that increasing $F$ has the same negative effect on diversity that was reported by Axelrod. The larger the number of features, the more likely it is that two agents will share the required number of features for influence to become possible. However, as $F$ increases further, the threshold number jumps to its next higher level, and the probability of influence drops sharply, followed by an increase as $F$ rises, until $F$ again hits the point at which the number of required features increases again, and so on.

To test our explanation of the effects of $F$ and $\tau$, we calculated the probability that two randomly chosen opinion vectors have sufficient overlap to have a positive probability of influence for a given threshold $\tau$, based on the binominal distribution. The probability that two agents agree on any given feature is $1/Q$, the probability that two independently drawn uniformly distributed random integers from the range [0..$Q$-1] obtain the same value. The probability that agreement occurs on exactly $k$ out of $F$ features, $p(F,k)$, is obtained from the probability density function of the binomial distribution. Technically,

$$p(F,k) = \binom{F}{k}\left(\frac{1}{Q}\right)^{k}\left(1-\frac{1}{Q}\right)^{(F-k)} \qquad (5)$$

The probability of influence being possible, $p^{*}(F,\tau)$, is obtained as the probability that agreement occurs on at least the threshold number of features, $f^{*}(\tau)$, hence

$$p^{*}(F,\tau) = \sum_{k=f^{*}(\tau)}^{F} p(F,k), \text{ where } f^{*}(\tau) = Min\{l \in \mathrm{N} \mid l \geq (1-\tau)F\} \qquad (6)$$

The function is graphed in Figure 4. The analytical results confirm the computational results in Figure 3. As $\tau$ approaches zero, the probability of influence increases and diversity declines accordingly in $F$. However, for a given threshold $\tau$, there is a secular decline in the probability of influence within those segments of the $F$-space for which the threshold number of features with identical traits remains constant.

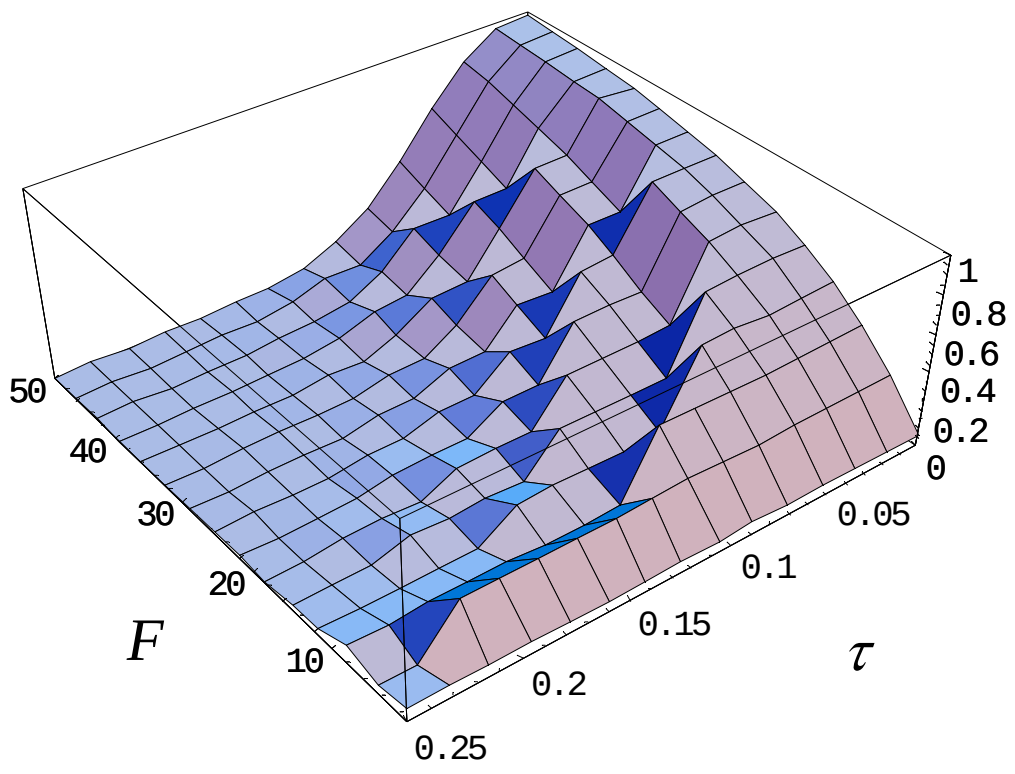

***Figure 4*. Probability of positive overlap between agents with random traits, as a function of *F* and *τ*, *F*=1..50, *Q*=15, *r*=0).**

In sum, these results suggest that the fragility of Axelrod's results reflect hidden assumptions in his model that attenuated the effects of homophily. His linear probability function implies a zero threshold of similarity. With zero thresholds, a low rate of mutation or a single metric feature is sufficient for diversity to collapse. However, as the threshold of similarity increases, diversity turns out to be much more robust than the original Axelrod assumption implies. In particular, we find that higher thresholds can reverse the effect of a larger number of cultural features. With a sufficiently high threshold, diversity now increases with the number of features. These results thus demonstrate the importance of homophily in explaining how local convergence can sustain global diversity.

## 3.4 Effects of Resistance to Deviants

In this section we report our test of the effects of multiplex influence on cultural diversity in Axelrod's framework. We began by testing whether multiplex influence can avoid the collapse of diversity that cultural mutation caused when the interaction threshold is zero. We replicated the experiment of figure 1, but replaced Axelrod's original assumption of dyadic influence with our implementation of multiplex influence. Figure 5 compares the average level of diversity across the two influence conditions, based on 20 replications per condition.

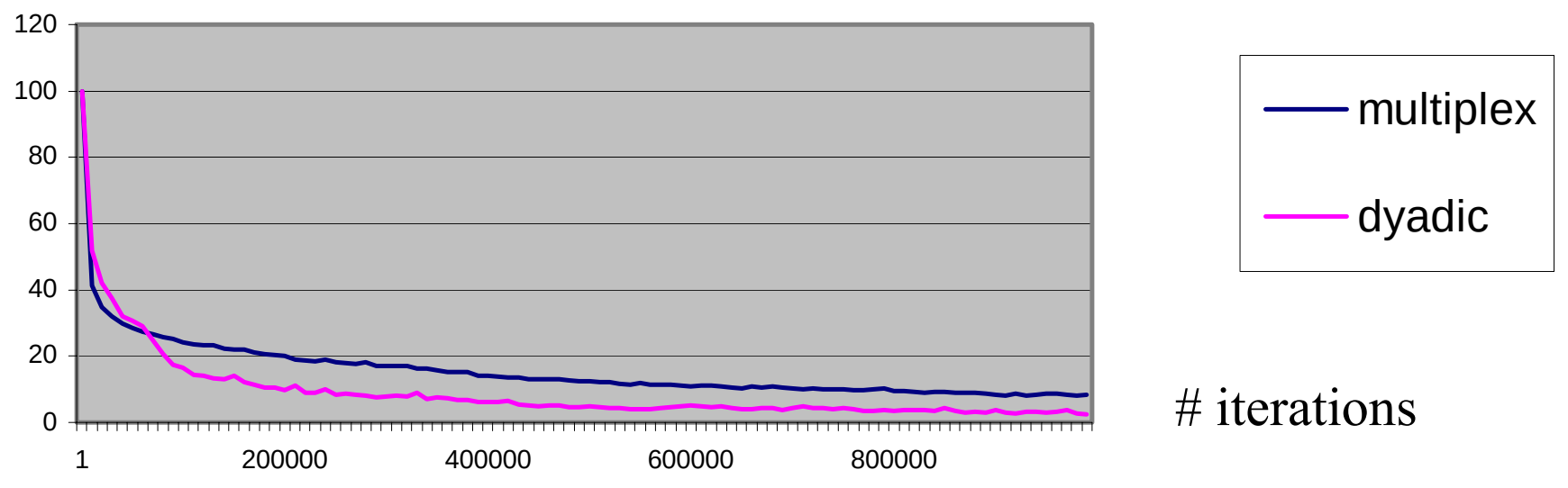

**Figure 5. Effect of dyadic vs. multiplex interaction on change of average number of cultural regions with mutation (*N*=100, *k*=4, *F*= 5, *Q*=15, $\tau$ = 0, *r*=0.001), based on 20 replications.**

Figure 5 confirms our expectation that multiplex influence greatly increases the robustness of cultural diversity against random mutations. With dyadic influence, average diversity after 1.000.000 iterations has declined to a level of about 2.4, compared to 8.3 cultural regions with multiplex influence. But this result may be attributed to two different explanations. We have argued that multiplex interaction makes cultural regions more robust. This suggests that once a cultural region has established, it remains stable for a longer period of time as it would under dyadic influence. But there is another possibility: under multiplex influence, diversity decays to monoculture just as much as it does for dyadic influence, but the system is more sensitive to mutations, that is: occasional outbursts of "cultural innovation" spread wider and create temporarily more diversity than they do under dyadic influence. In the average across multiple replications, the result would also be higher diversity under multiplex interaction than under dyadic interaction, but in any single run the most likely state after some time is monoculture both under multiplex and under dyadic influence. To exclude this possibility, we inspected single realizations simulated under the same conditions that we used for figure 5 (*N*=100, *k*=4, *F*= 5, *Q*=15, $\tau$ = 0, *r*=0.001). Figure 6 shows typical example runs.

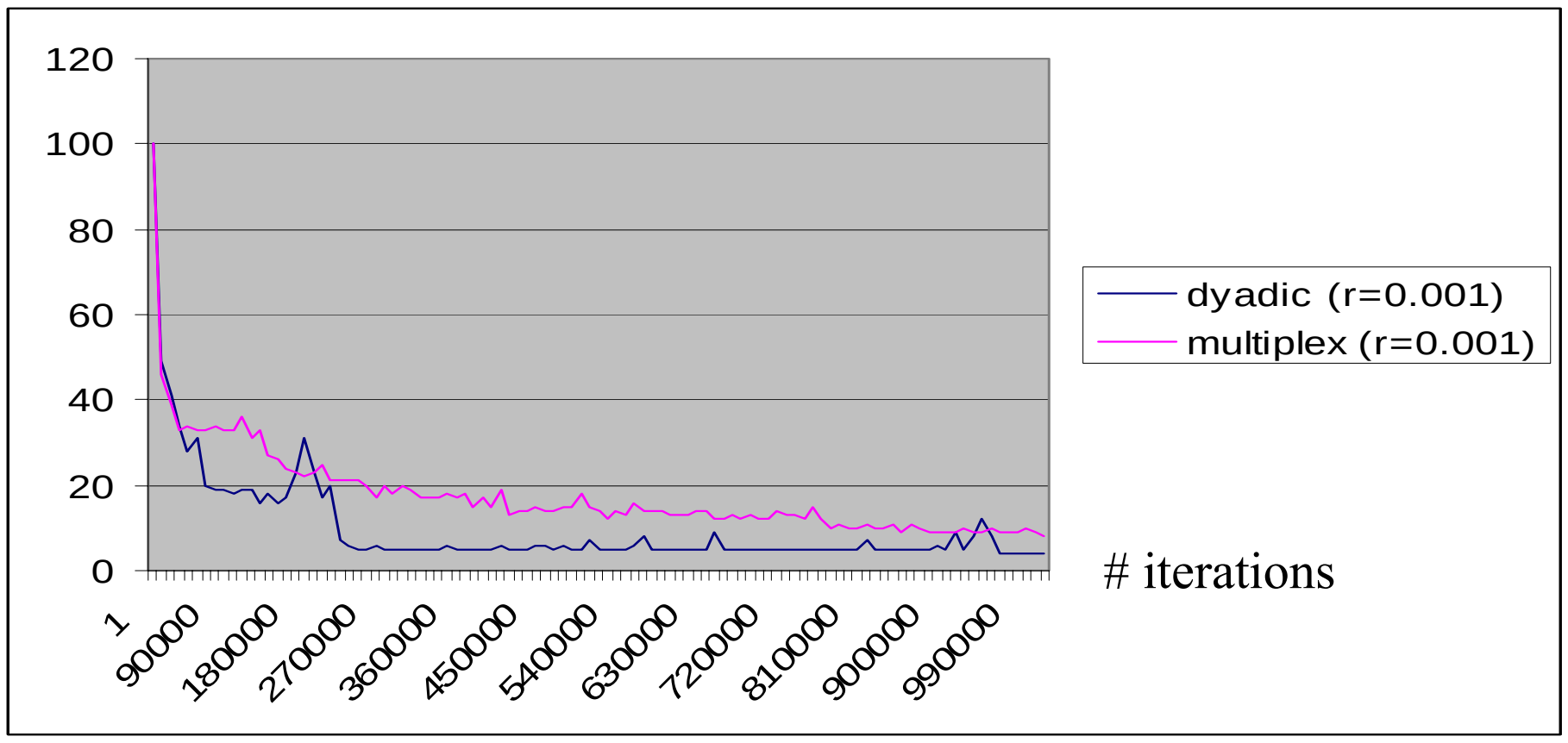

**Figure 6. Change of the number of cultural regions for single realizations under dyadic and multiplex interaction, with mutation (*N*=100, *k*=4, *F*=$F_n$=5, *Q*=15, $\tau$ = 0, *r*=0.001).**

Figure 6 indicates that simultaneous influence from all influential neighbors sustains the robustness of cultural regions. Not only is for *r*=0.001 the number of cultural regions after about 240.000 iterations in almost all iterations higher if influence is multiplex, but fluctuations in cultural diversity that are triggered by mutations are also clearly much larger under dyadic influence. While this result shows how multiplex influence stabilizes cultural regions, we can still not be sure that the effect is persistent. Moreover, the stabilizing effect of multiplex influence may also vanish when there is a larger level of random mutation. To test for both possibilities, we extended the experiment to 10.000.000 iterations (ten times as long) and varied the level of random mutation from *r*=0.001 to *r*=0.005 in steps of 0.001. For reliability, we computed for each condition the average number of cultural regions across 50 independent replications. Figure 7 charts the average number of cultural regions after $10^7$ iterations.

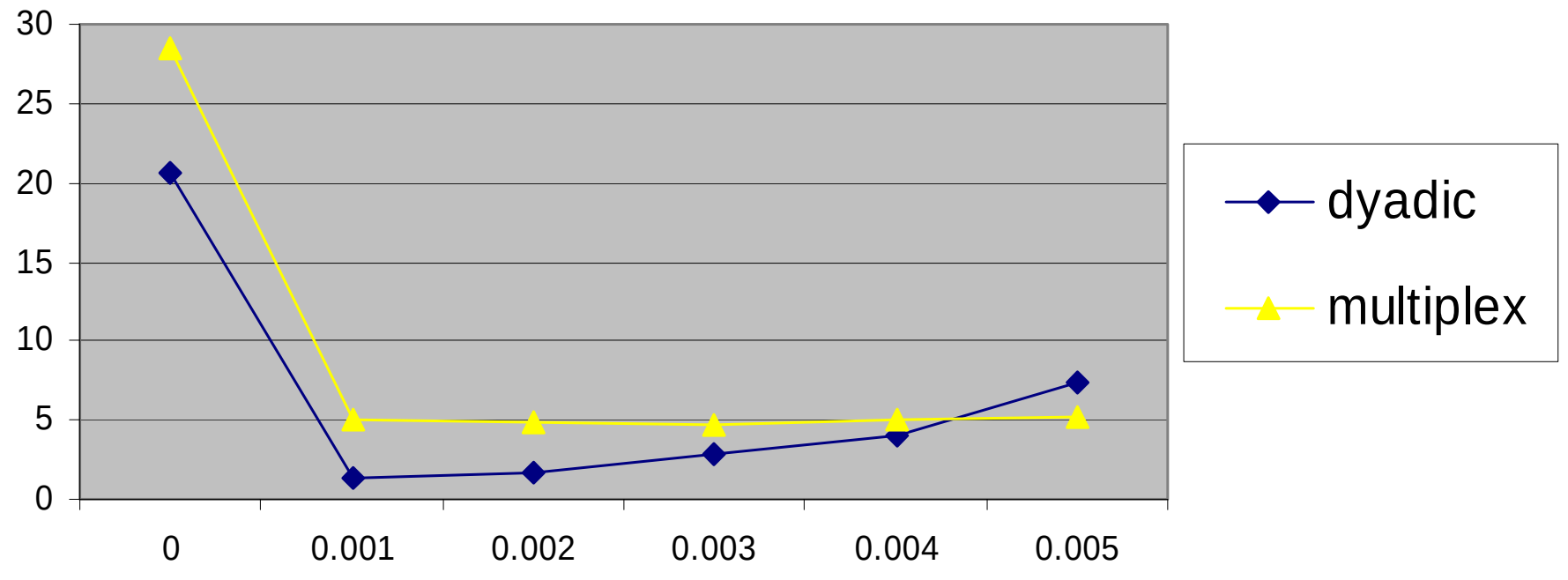

***Figure* 7. Average number of cultural regions in endstate (after maximally $10^7$ iterations) at six different levels of mutation (*N*=100, *F*= 5, *Q*=15, $\tau$ = 0, *r*=0..*r*=0.005), for multiplex and dyadic interaction, averages based on 50 replications per condition.**

The figure confirms that multiplex influence stabilizes cultural diversity against increasing levels of cultural mutation, and that the effect is persistent even after 10.000.000 iterations.

Figure 7 also reveals another difference between dyadic and multiplex influence. Average diversity increases with the level of mutation when influence is dyadic but not when it is multiplex. The explanation is the increase in the number "fads" precipitated by higher mutation under dyadic influence, as illustrated by Figure 6. When the population moves from one monocultural state into another one, there is temporarily a high level of diversity in the transition phase where the diffusion of the cultural innovation creates many different cultural profiles that are subsequently eliminated by the homogenizing effects of social influence. As Figure 7 shows, multiplex influence makes a population much more resistant to the outbreak of fads. The consequence is that monocultures do not arise (at least not within the time horizon of this experiment) under multiplex influence even at mutation rates as high as *r*=0.05. Under dyadic influence, the result is monoculture punctuated by brief outbursts of high diversity when the mutation rate is this high.

## 4. Discussion

Axelrod (1997) has proposed the combination of homophily and social influence as explanation of the stability of cultural diversity. However, subsequent work by Klemm and others has shown that stable diversity collapses when only small levels of cultural mutation are added to the model. We have shown that this fragility of Axelrod's result may in turn be an artifact of his overly conservative implementation of the assumption of homophily. We added interaction thresholds and multiplex social influence as two plausible modifications to the model. We showed how both modifications can independently restore diversity even under considerable levels of cultural mutation. Our analyses also point to new substantive implications of our modified cultural dissemination model. While Axelrod argued that more cultural features reduce diversity, we showed that with interaction thresholds, more features may increase diversity in a certain region of the parameter space.

Our analysis depends on a range of simplifications that point to directions for future research. Earlier studies of social influence (French 1956; Abelson 1964) point to an assumption in Axelrod's model that is problematic. Axelrod assumes that all cultural features are nominal, like religion and language (hence represented as nominal scales). With nominal features, people are either identical or different, there are no shades of gray. However, it is immediately apparent that cultures can also be distinguished by metric features with *degrees* of similarity, such as the normative age of marriage or the enthusiasm for jazz. Even religion and language can form nested hierarchies in which some classes are closer to others (e.g., Congregationalists are closer to Unitarians than to Islamic Fundamentalists). In a previous study (Flache and Macy 2006b) we found that local convergence does not lead to global diversity in Axelrod's model if even a single cultural dimension is metric, no matter how many dimensions are nominal. Cultural homogeneity is then the ineluctable outcome, even when we assume away random mutation. Future work needs to explore whether both of the solutions we propose in this paper also generalize to the case of metric features, and to settings where diversity is threatened by a combination of metric features and cultural mutation.

Another simplification concerns the model of the elementary mechanisms. In particular, previous studies of cultural dissemination including the present paper have largely neglected the negative side of homophily and social influence – xenophobia and differentiation. Some exceptions are recent studies by Mark (2003) and Macy et al (2003) and Flache and Macy (2006a) that allow for "negativity." Negativity may explain how agents remain different even when social interaction between them remains possible, but this has yet to be shown within the modeling framework of Axelrod's cultural dissemination model. We believe this is another promising avenue for future work.

References

Abelson, R.P. 1964. Mathematical Models of the Distribution of Attitudes Under Controversy. Pp. 142-160 in: Frederiksen, N. & Gulliken, H. (eds.): *Contributions to mathematical psychology.* Holt, Rinehart & Winston, New York.

Axelrod, R. 1997. The Dissemination of Culture: A Model with Local Convergence and Global Polarization. *Journal of Conflict Resolution* 41:203-226.

Carley, K. 1991. A theory of group stability. *American Sociological Review* 56:331-354.

Deffuant, G., Huet, S. & F. Amblard. 2005. An Individual-Based Model of Innovation Diffusion Mixing Social Value and Individual Benefit. *American Journal of Sociology* 110.4.1009-1041.

Festinger, L, S. Schachter, K. Back. 1950. *Social Pressures in Informal Groups*. New York, NY: Harper.

Flache, A. and M. W. Macy. 2006a. "Why More Contact May Increase Cultural Polarization." presented at Session Mathematical Sociology I at 101st Annual Meeting of the Ameri-can Sociological Association. August 14 2006. Montreal, Reprint published at www.arXiv.org > physics/0604196.

Flache, A., M. W. Macy. 2006b. "What Sustains Stable Cultural Diversity and What Undermines It? Axelrod and Beyond." presented at World Congress of Social Simulation 2006. August 24 2006. Kyoto, Japan (with K. Takács). Reprint published at www.arXiv.org > physics/0604201.

French, J. R. P. (1956) A formal theory of social power. *Psychological Review* 63: 181 –194.

Friedkin, N. E. and Johnsen, E. C. (1999): Social Influence Networks and Opinion Change. *Advances in Group Processes,* 16: 1-29.

Friedman, T.L. 2005. The World Is Flat: A Brief History of the Twenty-first Century. New York: Farrar, Straus & Giroux.

Greig, J. M. 2002. The End of Geography? Globalization, Communications, and Culture in the International System. *Journal of Conflict Resolution* 46.2:225-243.

Kitts, J., M.W. Macy, A. Flache, 1999. Structural Learning: Conformity and Social Control in Task-Oriented Groups. *Computational and Mathematical Organization Theory* 5:2:129-145.

Klemm, K., Eguiluz, V.M., Toral, R. & San Miguel, M. 2003a. Global culture: A noise induced transition in finite systems. *Physical Review E* 67, 045101(R).

Klemm, K., Eguiluz, V.M., Toral, R. & San Miguel, M. 2003b. Non-equilibrium transitions in complex networks: A model of social interaction. *Physical Review E* 67, 026120 (R).

Macy, M.W., J. Kitts, A. Flache, S. Benard. 2003. Polarization in Dynamic Networks: A Hopfield model of emergent structure. Pp 162-173 in: Ronald Breiger, Kathleen Carley and Philippa Pattison (eds). *Dynamic Social Network Modelling and Analysis: Workshop Summary and Papers*. Washington: The National Academies Press.

Mark, N. 1998. Beyond individual differences: Social differentiation from first principles. *American Sociological Review* 63:309-330.

Mark, N. 2003. Culture and Competition: Homophily and Distancing Explanations for Cultural Niches. *American Sociological Review* 68:319-345.

McPherson, M., L. Smith-Lovin, J.M. Cook. 2001. Homophily in social networks. *Annual Review of Sociology* 27:415-444.

Nowak, M. A., Szamrej, J. and Latané, B. 1990. "From Private Attitude to Public Opinion: a Dynamic Theory of Social Impact." *Psychological Review* 97(3):362-76.

Parisi, D., Cecconi, F. and Natale, F. 2003. Cultural Change in Spatial Environments. The Role of Cultural Assimilation and Internal Changes in Cultures. *Journal of Conflict Resolution* 47.2:163-179.

Shibanai, Y., Yasuno, S. and Ishiguro, I. 2001. Effects of global information feedback on diversity. Extensions to Axelrod's Adaptive Culture Model. *Journal of Conflict Resolution* 45.1:80-96.